\documentstyle[aas2pp4]{article}

\slugcomment{}

\lefthead{Lara et al.}
\righthead{The Radio-Optical Jet in NGC~3862}

\begin{document}

\title{The Radio-Optical Jet in NGC~3862 from Parsec \\
to Sub-Kiloparsec Scales}

\author{L.  Lara\altaffilmark{1}, L.  Feretti\altaffilmark{2}, G.
Giovannini\altaffilmark{2,3}, S.  Baum\altaffilmark{4}, W.D.
Cotton\altaffilmark{5}, C.P.  O'Dea\altaffilmark{4} and
T. Venturi\altaffilmark{2}}

\altaffiltext{1}{Instituto de Astrof\'{\i}sica de Andaluc\'{\i}a, CSIC, Apdo. 3004, 18080 Granada, Spain} 
\altaffiltext{2}{Istituto di Radioastronomia, CNR, Via Gobetti 101, I-40129 Bologna, Italy} 
\altaffiltext{3}{Dipartimento di Fisica, Universit\'a di Bologna, via B. Pichat 6/2, I-40127 Bologna, Italy}
\altaffiltext{4}{Space Telescope Science Institute, Baltimore, MD, USA}
\altaffiltext{5}{National Radio Astronomy Observatory, 520 Edgemont Road, Charlottesville, VA 22903-2475, USA}

\begin{abstract}

We  have observed the  radio source in NGC~3862 (3C264) simultaneously
with the EVN and the  MERLIN arrays, obtaining  detailed images of its
radio  structure from parsec  to sub-kiloparsec scales.  3C264 shows a
one-sided jet, with evident variations in its morphological properties
with distance.  We have analyzed HST optical data of NGC~3862, finding
a one-to-one correspondence between radio  and optical features in the
jet.  The  radio to optical  spectral  index is approximately constant
along  the jet.  Synchrotron  appears as the  most plausible mechanism
for the observed jet emission at radio and optical wavelengths.  Local
particle  reacceleration   or,   alternatively,   deviation  from  the
condition of equipartition are plausible mechanisms to explain the low
synchrotron cooling and constant spectral index in the jet of 3C264. 

\end{abstract}


\section{Introduction}

The interstellar medium plays a crucial role  in the spatial evolution
of extragalactic   radio jets.   There  is   increasing evidence  that
interaction   with  the   surrounding   environment   causes  jets  in
Fanaroff-Riley  type I radio sources  (FR  I; \cite{fr}) to decelerate
from initially relativistic  to sub-relativistic bulk velocities,  and
that such interaction determines  the morphology of their  large scale
structures  (\cite{parma}).  Observations  of  FR I  radio sources  at
parsec scales made with very long baseline interferometry (VLBI) show,
in  most cases, one-sided core-jet  morphologies. This fact is thought
to be a  strong indication of the  existence of  relativistic flows in
the inner parts of FR I jets (\cite{giovannini}; \cite{lara}).  On the
other hand,  observations   on the  kiloparsec  scale show  structures
consisting of  a  core and  two jets with   brightness ratio  close to
unity,  thus indicating that     relativistic boosting of   the  radio
emission      is  negligible at   larger    distances   from the  core
(\cite{parma}).  Current models for  FR I radio sources postulate that
deceleration of the plasma  flow from relativistic to  subrelativistic
speeds occurs  within 10 -   3000 pc from the   active core 
(\cite{bicknell}; \cite{laing}).  Needless  to say
the importance  of studying in detail  the regions where  the velocity
transition is   supposed  to take  place in   FR I  jets  in order  to
understand  the mechanisms of  plasma deceleration.  However, detailed
radio images  of  the sub-kiloparsec regions  of  jets in  FR I  radio
sources are not common  in the literature, mainly due  to the lack  of
sensitive interferometric arrays at  these scales. This difficulty can
be  overcome via the combination of  different  arrays.  Moreover, the
few known sources where jets   are observed at wavelengths other  than
radio are extremely  interesting, since broadband studies can  largely
constrain the proposed physical models. 

In this  framework,  we present and discuss  here  the results derived
from simultaneous  observations with the   European VLBI Network (EVN)
and the Multi Element Radio  Linked Interferometer Network (MERLIN) of
the FR I radio source 3C264 (B 1142+198) at 5 GHz.  The combination of
both instruments provides the angular  resolution suitable to map with
great   detail   the radio     structure   of 3C264   from parsec   to
sub-kiloparsec scales.   The large scale   structure of 3C264  is well
known: it exhibits a head-tailed morphology with  a prominent core and
a wiggling jet extending towards the northeast, which ends in a blob of
emission  at    35''  from the  core   (\cite{gavazzi}; \cite{bridle};
\cite{baum88}).  There   is  some  evidence   of  emission   from  the
counter-jet at about 11'' southwest from the core (\cite{lara}).  Both
the  jet and possible counter-jet  are embedded in  a vast and diffuse
emission  which seems  to be  dragged along the  north, revealing  the
existence of   a dense  intergalactic    medium.  In fact,    3C264 is
optically identified   with  NGC~3862 ($m_{v}=13.67$;   $z=0.0216$),  a
bright  galaxy in  the  cluster Abell  1367.  It   contains a  compact
($\sim$2'') and possibly variable X-ray source embedded in the diffuse
X-ray  emission from the   hot halo  of the   cluster, in a   position
consistent  with  that of  the central 5   GHz  radio source component
(\cite{elvis}; \cite{prieto}). 

NGC~3862 is  peculiar because it harbors  an optical jet  in a position
angle  coincident with that  of  the radio  jet (\cite{crane}; Baum et
al. 1997).  The optical jet, observed with the HST, extends up to 2''.
There is also evidence  of a ``ring'' around  the core, with a  radius
between 0.75'' and 1''   (Baum et al.  1997,  hereafter  BOG97).   The
region inside this ring seems to be cleared out of dust by the
jet  or, most  probably,    by  some  other nuclear  related   process
(\cite{hutchings}).  Line-emitting  gas is also   found  in  the  core
region, although it is not spatially  correlated with the jet emission
(\cite{baum97}). 

We assume a Hubble  constant $H_{0}=75$ km  s$^{-1}$ Mpc$^{-1}$ and  a
deceleration parameter  $q_{0}=0.5$,  and define the  spectral  index,
$\alpha$, such that $S_{\nu}\propto\nu^{-\alpha}$. In 3C264, 1 mas
corresponds to 0.4 pc in linear distance.

\section{Observations and Data Analysis}

We observed 3C264 simultaneously  with the EVN and  MERLIN on 1995 May
22.  The observations  were  made at the  frequency of  5 GHz in  left
circular  polarization  (LCP)  at     the  VLBI antennas,   and   dual
polarization at the MERLIN antennas.  The bright radio sources 4C39.25
and 1144+402 were also observed  as fringe finders for the correlation
of the VLBI data, and as point source calibrators.

The  MERLIN  array  consisted  of  six  antennas   (Tabley, Cambridge,
Jodrell-MK2, Darnhall, Knockin  and Defford, all  located in England),
with  minimum   and  maximum baselines   of   130  k$\lambda$ and  2.2
M$\lambda$, respectively,  recording  a  bandwidth of 15    MHz.  Flux
density calibration of  the MERLIN  data was  performed with the  OLAF
package by  comparison of  OQ208 and  1144+402 with  the  primary flux
density calibrator 3C286 (see \cite{baum97} for a complete description
of MERLIN observations and data reduction). 

The  EVN  array consisted of    seven antennas: Effelsberg  (Germany),
Cambridge  and Jodrell-MK2  (U.K.), Medicina and  Noto (Italy), Onsala
(Sweden) and  Westerbork  (The Netherlands).  The  minimum and maximum
baseline separation were 3.7 and 38 M$\lambda$, respectively.  The EVN
stations used the   Mark-III recording system   (\cite{rogers}) with a
synthesized  bandwidth of 28  MHz  (mode B).  The  VLBI  data from the
different telescopes      were   correlated  in  absentee     at   the
Max-Planck-Institut  f\"{u}r    Radioastronomie  in Bonn    (Germany).
Cambridge did not yield VLBI fringes  due to technical problems during
the  recording of the    data.    The calibration of the    visibility
amplitudes  was performed with the  NRAO Astronomical Image Processing
System (AIPS). In addition, the  correction of the fringe rate through
the so-called global fringe fitting (\cite{schwab}) was applied. 

We used the Caltech Difmap  package (\cite{shepherd})
and  the  AIPS  package  for imaging the  EVN   and the  MERLIN  data,
respectively,  following   standard  self-calibration   and    mapping
procedures.  Once the separate maps from each  array were produced, we
proceeded  to combine  the self-calibrated EVN  and  MERLIN  data. The
different  calibration schemes applied to  each data set resulted in a
slight  misalignment between the MERLIN  and  EVN flux density scales.
Considering that  MERLIN  is sensitive to   extended emission which is
resolved  out by  the EVN, and  that the  flux  density calibration of
MERLIN  was made using a primary  flux density calibrator, we adjusted
the EVN flux density scale to fit the more reliable MERLIN scale.

Final maps from the  combined data set were  obtained with  the Difmap
package.  The EVN image,  which reproduces well   the data from longer
baselines, was taken as the    initial  source model in the    mapping
procedure.  In order to find a suitable  delta-component model able to
reproduce the visibility data at  all uv-spacings, we allowed CLEAN to
act  progressively on  more and  more further  regions along  the jet.
Data  with  large  uv-spacings  were  also weighted less  by applying
progressively stronger Gaussian uv-tapers in order to gain sensitivity
to more extended  emission   after  each self-calibration and    CLEAN
iteration.

\section{Results}

\subsection{The Jet from Parsec to Sub-kiloparsec Scales }

In Fig.\ref{fig1} we  present images of the radio jet in NGC~3862,  
ordered from highest (6.0 mas  $\times$ 3.6 mas), at top-left,
to lowest (62  mas $\times$ 52  mas) angular resolution, at top-right.
Details of  the  radio  structure  from scales  of several  parsecs to
scales of $\sim$0.5 kpc are evident. 

The EVN  map in Fig.\ref{fig1}a shows  an unresolved core and a smooth
one-sided   jet extending up  to 25  mas from the  core along position
angle (P.A.) $\sim 27\arcdeg$. From Fig.\ref{fig1}a it is difficult to
identify discrete components in the    innermost jet of 3C264.     The
comparison of  this map with  a previous  global VLBI  map made at the
same frequency (\cite{lara}) shows that,  at  the level of  resolution
provided by the EVN, there is no  indication of variability of the
structure
on  a time scale  of two years,  while  flux density variability might
exist within a 10\% level. 

Fig.\ref{fig1}a  shows hints of  emission at a distance  of 55 mas (22
pc) northeast  from the core. A  well defined  component is evident at
this position when MERLIN data  are used (Fig.\ref{fig1}b).  From this
component on, the jet appears narrow, well collimated and slightly
curved towards the east.  In fact, the jet position angle changes from
27\arcdeg\ in  the inner jet regions,  to  44\arcdeg\ at $\sim$250 mas
from the core.  This  collimated  region is distinguished  by  several
knots  and by a  slowly decaying surface brightness (Fig.\ref{fig1}c).
At 250 mas  (100 pc) from the core  there are important changes in the
jet characteristics: the opening angle of  the jet increases abruptly,
and  the    flow becomes   apparently   more  turbulent.   We  observe
filamentary emission, pronounced kinks  and wiggles which  suggest the
existence of   oscillations  in the   ridge  line  of    the jet,   or
alternatively, a  three dimensional  helical structure evolving  along
three  great  loops (Fig.\ref{fig1}c-d-e).  The    cone angle of   the
observed   oscillations    is   about   15$^{\circ}$.    This possible
oscillating or helical  structure  is also marginally visible   in the
region between 50 and 100 pc from the core. 

\placefigure{fig1}

The lower resolution maps (Fig.\ref{fig1}e-f) show a  sharp cut in the
jet  surface brightness at   a distance  of  0.8  asec from  the  core
($\sim$320  pc), and a deflection  of  the ridge line. This deflection
could be related with an abrupt change in the external medium, perhaps
associated with the transition of the jet from the central region, 
cleared out of dust, to the region outside the ring    (\cite{baum97};
\cite{hutchings}).  With the  sensitivity of the  observing array, the
jet emission is detected up to a distance of 1.2 asec. 

The MERLIN array, without Cambridge, recorded both left and right hand
circular polarizations. We  made a  map of  the polarized emission  of
3C264, although with  lower  angular resolution (156   mas $\times$100
mas,  P.A. -53\arcdeg) if compared with  the total intensity maps, due
to  the lack of the longer   MERLIN baselines.  Although some problems
with  the calibration  could not  be definitely solved  (the rms noise
level of the final  image was 0.12 mJy/beam), significant polarization
is detected  in a small region (2  beam area) at a  distance of 170 pc
from the core.  In this  region the  projected magnetic field  appears
parallel to  the jet  axis and the  average degree  of polarization is
16\%. 

There   is no  evidence  of  a  counter-jet in   any  of the  maps in
Fig.\ref{fig1}. Lower limits of  the   jet to counter-jet   brightness
ratio (R)  of  3C264 obtained with   different angular  resolutions at
different    positions     along    the  jet     are    presented   in
Table~\ref{tab1}.  To compute R, we assume  that the emission from the
counter-jet is lower than  or equal to 3 times  the rms noise level of
the image, measured over large empty regions of the maps. 

\placetable{tab1}

In summary, four  distinct parts can   be distinguished in  the jet of
3C264, indicating that it does not expand  at a constant rate as would
a steady  free jet:  {\em  i)}   the strong  compact core  and  smooth
innermost jet   region (0  -  10 pc);  {\em   ii)} a  narrow  and well
collimated  region which  extends up to  110 pc;  {\em iii)}  a region
where the  jet widens,  showing kinks and  filaments which  suggest an
oscillating or   a helical jet. This  region  has a projected magnetic
field predominantly parallel to the jet axis (100 - 300 pc); {\em iv)}
a  faint and narrow region  appearing  after a  strong jet deflection,
possibly due to interaction with the ring (300 - 430 pc).

\subsection{Comparison of Radio and Optical Emission}

As already noted in BOG97, the optical and radio emission from the jet
in NGC~3862 show a remarkably similar structure. A detailed comparison of
HST and radio images  faces up to the problem  of how reliable is  the
subtraction of the surface  brightness from the elliptical host galaxy
in  order to  highlight the optical   jet.   Fig.4 in  BOG97 shows the
optical jet of NGC~3862,  with   deep negative contours surrounding   the
core. Brightness profiles taken  at several position angles around the
core suggest that the subtracted elliptical galaxy model overestimates
the  optical  emission in the  inner $\sim$70  pc, thus  producing the
observed negative contours.   This region  will  not be  considered in
this Section. We also  note that a factor of  2 must be applied to the
flux density scale in the optical image  of NGC~3862 in BOG97, because 
a change in the software used to combine the images for cosmic ray removal 
caused confusion over whether the image header was updated with the sum of 
the exposure times.

\placefigure{fig2}

To compare the radio and  the HST images, both  were convolved with  a
circular Gaussian beam of 45 mas.  The  images were then registered by
the peak of brightness of the core in the two maps. Opacity effects on
the position of  the core are assumed  to be negligible  compared with
the actual  resolution of 45 mas.  We  find that, beyond a distance of
100 pc from the core, there is good correspondence of radio and optical
features in the map.  Fig.\ref{fig2}a shows brightness profiles of the
radio and optical  flux density along  the ridge line  of the jet. The
optical emission has been multiplied by  a constant factor for display
purposes.  There are five distinct peaks at radio wavelengths and four
peaks    at  optical  wavelengths.  Peak   positions    in the optical
correspond to peak  positions in the radio,  except for the first knot
at $\sim$90 pc  from the core, which  seems to be shifted  outwards at
optical wavelengths. We  however note   that this  shift could  be   a
consequence  of the elliptical  image subtraction in the inner regions
of NGC~3862. 

The radio-to-optical spectral  index  along the jet, obtained   with a
resolution of 45  mas,  is  displayed  in Fig.\ref{fig2}b.   It  shows
slight variations around a  mean value of  0.63, very similar  to that
found by Biretta et al.  (1991) in the jet of M87.  We estimate errors
of the  order of 0.02  (corresponding to 1~$\sigma$, including the rms
noise  level  of each image plus   a 15\% error  in  the  flux density
scale).  The  knots  show  a  steeper  spectrum   than  the
inter-knot  regions, with differences in  the  spectral index of about
0.05,   thus marginally significant  if  compared with the measurement
uncertainties.  However,  Hutchings  et al.   (1998)  report from  HST
optical  spectroscopy  that  the    knots  are  brighter  at   shorter
wavelengths.  The steeper spectrum we  find in the knots results  from
the less contrast observed  between jet-knot brightness in the optical
than in the radio (Fig.\ref{fig2}a), a feature which might be real but
also consequence of the subtraction of the background galaxy emission.

We have measured the  flux density of the  jet in NGC~3862 using  present
radio observations at 5 GHz, recent 1.6 GHz MERLIN observations (to be
published elsewhere), optical data in BOG97, data available in the HST
public archive, and data from the literature (Table~\ref{tab2}).  Flux
density  of the radio jet has  been estimated in  AIPS measuring first
the total  flux  density over  the  region  where optical emission  is
observed, including the core, to  which the integrated flux density of
a Gaussian component which fits the core emission is later subtracted.
The HST optical data  have been  analyzed using  IRAF: we  removed the
cosmic-ray hits with  the task CRREJ, and constructed   a set of  four
images by rotating the  original one 0\arcdeg,  90\arcdeg, 180\arcdeg\
and 270\arcdeg,  with the core   peak  of brightness  at the  rotation
center. The median of  these four images,  a good approximation to the
galaxy core and  background emission, was subtracted from the original
image to isolate the jet. 

The    broadband  spectrum of  the jet is  displayed in
Fig.\ref{fig3}.  The radio spectral index  ($\alpha_{r}$) is 0.46, and
the  optical  spectral index ($\alpha_{opt}$)  is  1.34.  Crane et al.
(1993) report  a similar optical  spectral index for  the jet 
($\alpha_{opt}$=1.4)   from  HST observations   at  two   wavelengths,
although their flux  density values are much higher  (a factor of 2.5)
than the values  derived  by us.  Flux  density variability  seems  an
unlikely reason  for this  large discrepancy, since   only a very  low
level of  variability is observed  at radio wavelengths.  For the sake
of consistency,  Crane et al.  (1993) data  have not been  used in the
previous  estimation of the optical  spectral  index, and will not  be
used for further discussion. 

The Rosat X-ray flux density (\cite{prieto}) is also plotted in
Fig.\ref{fig3}.  This measurement is clearly an upper limit due to the
low  angular  resolution (25'') of the   X-ray observations.  In fact,
Prieto finds that no single power law can be fit  to the compact X-ray
emission, indicating that it might be contaminated by the contribution
from a thermal component. 

\placetable{tab2}
\placefigure{fig3}

\section{Discussion}

\subsection{Jet Dynamics}

>From Fig.\ref{fig1}  it is  evident that the   jet of 3C264  does  not
expand at   a  constant rate. It appears   narrow  and well collimated
during the first  100 pc, and  widens further  on, apparently becoming
turbulent.  Deceleration between parsec and kiloparsec scales in 3C264
was suggested by  Lara et al. (1997) from jet to counter-jet brightness
ratio arguments, assuming  an intrinsic symmetric structure. Moreover,
BOG97  deduced an important jet  deceleration  between 300 and 400  pc
from the core, modeling the emissivity of an adiabatic expanding jet.
Deceleration was then explained through a change  of properties in the
external medium   associated with the   existence of the circumnuclear
ring, and an initial  velocity of 0.98c was required  in order to make
compatible the observed jet  to  counterjet brightness ratio with  the
adiabatic model predictions. 

The very high  resolution of the present  radio observations allows us
to  study the jet dynamics from  a few parsecs  up  to 300 pc from the
core.   From similar arguments   to  those  discussed in BOG97,   i.e.
assuming  adiabatic   expansion   and   a  magnetic   field   oriented
predominantly parallel to  the jet  axis, we obtain  that  in order to
compensate for the dramatic expansion of the  jet opening angle at 100
pc  and  keep  the predicted  surface brightness    comparable to that
observed, the jet has to decelerate in the  region between 100 and 200
pc  from   0.98c  to 0.85c.    However, this   result faces   two main
difficulties: first, the  observed  degree of polarization implies  an
important contribution of  random field in  3C264, so that assuming  a
parallel and   uniform magnetic field  may  be not  a good assumption.
Second,  if the  jet had   an  oscillating or helical structure,   our
estimate  of  the  evolution of   the jet  width would  be misleading,
seriously affecting the  prediction of the  adiabatic expanding model.
In  conclusion, although  deceleration  between parsec and  kiloparsec
scales  is rather  plausible in  3C264 (\cite{lara}),  the lack  of an
observed counter-jet  at  sub-kiloparsec  scales and  the difficulties
inherent to the adiabatic model do not allow us to precisely determine
if deceleration is taking place at these scales. The dramatic widening
of the jet  at  100 pc  from the core   could also be  due  to the jet
entering  a   region of decreased external   pressure,  rather than to
deceleration of the jet. 

\subsection{Estimation of Physical Parameters}

We obtained surface brightness profiles perpendicular  to the jet main
axis at several  positions along the  jet,  and then measured the  jet
FWHM  and the  surface   brightness at  the ridge  line  from Gaussian
fits.  The deconvolution of jet  width and surface brightness was done
following Appendix A in Killeen et al.  (1986).   We used the standard
formulae of synchrotron radiation (e.g. \cite{miley}) to calculate the
minimum energy  density    ($u_{me}$)  at these   positions    and the
corresponding   effective     magnetic  field   $B_{me}$,   which   is
approximately the equipartition field.  The  total pressure is assumed
to be  that of  equipartition  between particles  and  magnetic field,
$P_{eq}=0.62~u_{me}$.  The effects of Doppler boosting were taken into
account assuming a  constant jet velocity of  0.98c up to 300 pc  from
the core (BOG97).  In addition, the following assumptions were made in
the calculations: {\em i)} the magnetic field is assumed to be random;
{\em ii)} the  energy of particles  is equally stored  in  the form of
relativistic electrons and heavy particles; {\em iii)} lower and upper
frequency cutoffs  were set to 10  MHz and 10$^{5}$ GHz, respectively;
{\em iv)} the spectral index of the jet  is 0.5 (Fig.~\ref{fig3}); and
{\em v)} the  line-of-sight depth is  equal to the deconvolved FWHM of
the jet corrected for projection effects.  We assume that the jet axis
forms  an angle of 50\arcdeg\ with  respect  to the observer's line of
sight (such jet  orientation was taken  from BOG97 and is in agreement
with the  constraints  in Lara   et al.  (1997)  derived from   jet to
counter-jet brightness ratio and core prominence).  An upper frequency
cutoff of  10$^{5}$  GHz was  assumed  in  order  to allow synchrotron
emission at optical  wavelengths  (see also Section 4.3.2).   However,
for  the given  spectral  index, the results   on  the minimum energy,
pressure and magnetic field are  not very sensitive to this parameter.
The results obtained  at different positions  along the jet are listed
in Table~\ref{tab3} and displayed in Fig.\ref{fig4},  as a function of
the filling factor $\eta$. 

\placetable{tab3}
\placefigure{fig4}

The dependences of the magnetic field on the jet width ($\Phi$) and on
the distance from the nucleus  ($D$) could be  fit with similar  power
laws  of the form $B_{me}  \propto  \Phi^{-0.8}$ (Fig.\ref{fig4}c) and
$B_{me}   \propto   D^{-0.8}$   (Fig.\ref{fig4}d).   In  an  adiabatic
expanding jet, the expected trends of the magnetic  field with the jet
width   are $B_{\perp} \propto  \Phi^{-1}$  and $B_{\parallel} \propto
\Phi^{-2}$ (BOG97  and references therein).  Identical expected trends
are  derived for the magnetic  field with  the  distance from the core
(\cite{hughes89}).  We  obtain  dependences closer to $\Phi^{-1}$  and
$D^{-1}$,  appropriate   for a jet   with  an uniform   magnetic field
perpendicular to the jet  direction.  However, only  in a small region
of  3C264 we observe a  projected magnetic field  parallel  to the jet
main  axis with a degree  of polarization of  16\%, which implies that
the ordered magnetic field is 47\% of  total in this region, being the
rest   oriented randomly (if Faraday  depolarization   is ignored at 5
GHz).   If the jet  of 3C264  were purely  adiabatic, we would  expect
trends of the  magnetic  field with  slopes between   -1 and  -2  (see
Section 4.3.2 for further discussion).

\subsection{Mechanisms for Optical Emission in Jets}

There   is an  increasing  number  of radio   jets  for which  optical
counterparts have been observed: 3C273, M87, PKS~0521-36, 3C66B, 3C264
(see \cite{sparks94} for   a review), 3C78 (\cite{sparks95}),   3C200,
3C346   (\cite{koff}),   3C212,       3C245  (\cite{ridgway}),   3C371
(\cite{nilsson}),  and  3C15   (\cite{martel}).  The  understanding of
optical emission spatially coincident  with synchrotron radio emission
in jets is an open question.  In principle, if the radiation mechanism
at optical wavelengths were synchrotron, we would not expect to detect
optical emission at  large   distances from the  core  since radiation
losses   of electrons emitting at visible   wavelengths  are orders of
magnitude larger than those of electrons emitting at radio wavelengths
(\cite{felten}). However,  this is not the case,  and  optical jets of
kiloparsec lengths exist.   Several  scenarios have  been  proposed to
solve   this discrepancy,  mostly based on   observations  of the well
studied M87 radio source: {\em i)} First-order Fermi reacceleration of
electrons  in the knots  of a jet.  {\em  ii)} The existence of a high
field boundary layer surrounding a  low magnetic field channel through
which the    relativistic  electrons are  conducted,  with  negligible
synchrotron losses, to the place where they cross the high field shell
and radiate (\cite{owen}).  {\em  iii)} Local reacceleration processes
occurring  all along  the  jet, which  maintain roughly  constant  the
maximum cutoff energy  of  the electron population.  Spectral  and brightness
variations are produced  by changes in   the magnetic field  strength,
which dominates the appearance of the jet (\cite{meisenheimer}).  {\em
iv)} Magnetic  field  below equipartition  and adiabatic fluctuations,
combined with Doppler beaming effects (\cite{heinz}). 

Of course, different  explanations  for optical emission aligned  with
radio jets  are  possible    if  different emission   mechanisms   are
invoked. Besides  synchrotron, radiation  at optical wavelengths could
be produced  {\em i)}  by inverse Compton   scattering of  photons  by
relativistic electrons in the jet, {\em  ii)} by Thomson scattering of
anisotropic   light emitted  from   the core,   {\em  iii)} by thermal
bremsstrahlung   emission   from clumped hot  gas,    or {\em  iv)} in
jet-induced  star-formation  regions (see   Daly  1992 for  a detailed
discussion).   In NGC~3862, the  radio    and the optical emission   are
spatially coincident, with peaks  in the radio corresponding  to peaks
in the optical, except the first knot  which, as noted in Section~3.2,
could  be affected by the  galaxy  subtraction.  This implies that the
radiation from  both parts of the spectrum  answer in  the same way to
the  local variations in the   physical conditions.  This fact renders
rather unplausible for  the optical radiation  any mechanism which  is
not directly related to the  synchrotron radiation observed at radio
wavelengths.   In addition, the thrust of  the relativistic flow might
have swept  any thermal gas in the  region occupied by the jet itself,
leaving  Thomson or  bremsstrahlung processes   constrained to regions
well outside the jet axis, and  thus not spatially coincident with the
synchrotron  radiation. Note also that Hutchings et al. (1998) find no 
optical emission lines associated with the jet. 

In the next sections we   will concentrate on two possible  mechanisms
for  the optical  radiation  which   are  directly related   with  the
synchrotron processes  responsible for the radio emission: inverse
Compton scattering (IC) and synchrotron itself.

\subsubsection{Inverse Compton Scattering}

Following Blumenthal \& Tucker (1974) we  have derived the temperature
of an   external radiation field  described   by a Planck distribution
which is  IC scattered by the  population of relativistic electrons in
the jet: 
\begin{equation}
T=\left[\frac{4.05\times10^{18}}{4884^{\alpha}}\cdot \frac{a}{b}\cdot
B^{1+\alpha}\left(\frac{\nu_{opt}}{\nu_{rad}}\right)^{\alpha}
\frac{S_{ic}}{S_{s}} \right]^{\frac{1}{3+\alpha}},
\end{equation}
\noindent
where $\alpha(=0.5)$ is the spectral index, assumed constant along the
jet;  $a$    and  $b$ are  two   spectral   index dependent  functions
($\frac{a}{b}=0.02$ for $\alpha=0.5$);   $B$ is the average  effective
magnetic field, which we have obtained from minimum energy assumptions
($B\sim$0.7  mG in  the    region where strong  optical   radiation is
observed);   $\nu_{opt}$(=4.35$\times   10^{14}$       Hz)         and
$\nu_{rad}$($=5\times 10^{9}$  Hz)  are the  observing frequencies  of
optical and radio emission, respectively; $S_{ic}$ and $S_{s}$ are the
jet  flux densities   measured   at  optical and   radio  wavelengths,
respectively  (see  Table~\ref{tab2}).   From equation  1  we obtain a
black body temperature of 590 K, which, according to Wien displacement
law,  corresponds to  the maximum energy   emitted  at a frequency  of
$3.5\times10^{13}$ Hz   (9$\mu m$).   A hot    dusty  region could  be
responsible of such  radiation field in  the infrared.  The integrated
flux density derived at 60$\mu m$  from a region  with a radius of 400
pc  turns to  be of  the order of  $10^{4}$ Jy,  which is indeed  much
higher than  the observed 220   mJy at this  wavelength  (\cite{hes}).
Thus, we conclude that it is unlikely that the optical jet emission is
produced by IC scattering of an external radiation field. 

In a similar way, we can estimate the  ratio of the IC and synchrotron
radiation,  assuming this time that  the  synchrotron photons from the
jet    are  IC   scattered  by   the   relativistic  electrons    (see
\cite{hughes91}): 
\begin{eqnarray*}
\frac{I_{\nu,ic}}{I_{\nu,s}} &  =  &
\frac{16\pi g {\cal F}(\alpha)}{\sqrt{3}\left(\frac{3}{2\pi}\right)
^{\alpha} c_{5}^{(\alpha)}(\alpha)  } \cdot
\frac{4\pi\epsilon_{o} c \sigma_{T}}{e^{2}} \cdot
I_{\nu,s}~\nu_{s}^{2\alpha}~\ln\left(\frac{\nu_{u}}{\nu_{l}}\right)\\
  &  & \left(\frac{3\nu_{ic}}{4}\right)^{-\alpha} \cdot
\left(\frac{eB}{m_{e}}\right)^{-(\alpha+1)} \cdot
\left(\frac{{\cal D}}{1+z}\right)^{-(\alpha+3)}.
\end{eqnarray*}
In this formula, expressed in SI units, $g$ is a geometrical factor to
allow    for possible  anisotropy    of   the   photon  and   electron
distributions; ${\cal  F}$  and $c_{5}^{(\alpha)}$ are spectral  index
dependent functions (see reference for exact definition). In our case,
$\alpha   =       0.5$,          ${\cal  F}(\alpha)=0.455$         and
$c_{5}^{(\alpha)}(\alpha)=   2.945$;   $\sigma_{T}$   is   the Thomson
cross-section; $\epsilon_{o}$ is the permittivity  of free space;  $c$
is the velocity of light; $e$ is the charge of an electron; $\nu_{s}$,
$\nu_{ic}$, $\nu_{u}$  and $\nu_{l}$ are  the frequency of synchrotron
radiation (5 GHz),  IC scattered  radiation ($4.35\times10^{5}$  GHz),
and upper (100 GHz) and lower (0.01 GHz) limits to frequency spectrum,
respectively; $B$ is the magnetic field ($7\times 10^{-8}$ T); $m_{e}$
is the electron mass; ${\cal D}$ is the Doppler factor. 

We  obtain that
\begin{equation}
\frac{I_{\nu,ic}}{I_{\nu,s}}\simeq 2.2 \times 10^{-9}~g~{\cal D}^{-3.5},
\end{equation}
which has to be compared with an observed  ratio of $8\times 10^{-4}$.
If we assume  $g\simeq{\cal D}\simeq  1$,  then the  predicted optical
emission  is several orders of   magnitude too low making it  unlikely
that the  IC mechanism  is responsible  for the optical  emission.  IC
emission would be consistent with the data only for ${\cal D}=0.03$ if
$g\simeq 1$.  Such  a low Doppler  factor would  require the  jet (and
counter-jet) bulk  velocity  to   be  very close to    $c$  ($\beta\ge
0.9995$), which is  a very high velocity if  compared with  other well
studied  radio  jets, but in addition,  would  render 3C264 one of the
intrinsically brightest sources on   the sky, which contrast  with  the
level of emission from the lobes and with its  classification as an FR
I radio source.

In summary, we conclude that IC is not a plausible mechanism for the 
jet optical emission in NGC~3862.

\subsubsection{Synchrotron Emission}

Previous arguments  and  the   broadband  spectrum  in  Fig.\ref{fig3}
suggest synchrotron as the  most plausible mechanism of  emission from
radio to optical wavelengths in the  jet of NGC~3862.  In consequence,
we have  fit three different synchrotron spectral  aging models to the
broadband spectrum of  the jet: the  continuous injection  (CI) model,
the  Kardashev-Pacholczyk (KP) model, and  the Jaffe-Perola (JP) model
(\cite{kardashev};  \cite{pacho}; \cite{JP}).   These models assume an
initial power-law distribution of particles.  The CI model considers a
continuous injection of fresh particles in the distribution, while the
KP and JP models assume a  single ``one-shot'' injection of particles.
The JP model allows for isotropization of particles pitch-angle, while
the  KP   model   assumes  constant  pitch-angles.     Each model   is
characterized  by an injection spectral   index and a  break-frequency
above which  the   spectrum steepens.   We obtain   that  an injection
spectral  index   of 0.5 is valid   for  all three models,  with break
frequencies equal to  $4.3\times 10^{13}$ Hz, $2.2\times 10^{14}$  Hz,
and $4.3\times 10^{14}$ Hz for the  CI, KP and JP models, respectively
(Fig.\ref{fig5}).   The  KP model   provides  a  slightly  better  fit
($\chi^{2}$) than  the  CI or JP   models, although our data   are not
adequate to discern any significant differences in these models. 

\placefigure{fig5}

Optical synchrotron radiation  requires  high energy  electrons,  with
$\gamma$ of   the  order of  $10^{6}$,  which  imply short synchrotron
cooling times.  With  the magnetic field  derived from minimum  energy
assumptions ($B_{me}$ (Gauss)), and the upper frequency cutoff derived
from the KP model ($\nu_{b}$ (GHz)), we estimated the mean synchrotron
lifetime at several positions along the jet, $<t_{s}>$ (\cite{miley}): 
\begin{equation}
<t_{s}>(yr)=0.82~B_{me}^{\frac{1}{2}}
\left(B_{me}^{2}+B_{MW}^{2}\right)^{-1}
\left(1+z\right)^{\frac{-1}{2}} \nu_{b}^{\frac{-1}{2}}
\end{equation}
where  $B_{MW}=3.5\times10^{-6} (1+z)^2$   (Gauss) is  the  equivalent
magnetic  field of the  microwave  background. The result is that the
lifetime   increases smoothly  with the  distance  from  the core (see
Table~\ref{tab3}), reaching a maximum value of only 190 yr at 300 pc. 

Assuming a  lifetime   of  the order  of 2$\times$10$^{2}$    yr,  the
electrons must have been accelerated, at the most, at a distance of 60
pc   (150 mas) from the  location   where radiation is being produced.
Reacceleration  by  Fermi processes in  shocks  would then require the
existence of well defined knots  all along the jet  of 3C264, which we
do not   observe in our maps.   In  addition, it  is  known that Fermi
reacceleration finds difficulties in explaining the constancy of radio
to optical spectral  index in M87 (\cite{heinz}). Similar difficulties
would arise  for 3C264, which  also shows  a rather  uniform radio  to
optical spectral index.  The slight  steepening of the spectrum in the
brightness peaks (Fig.\ref{fig2}),  if real, would indicate that these
are regions    of  more   efficient energy  dissipation     at  higher
frequencies, rather than  places of particle reacceleration, and could
hardly be associated with the shocks required by the Fermi mechanism. 

If the wiggles  observed in 3C264  were  signs of a helical  structure
wrapped around an empty cone, then it would be  possible to invoke the
hollow channel model (\cite{owen})  to explain the lack of synchrotron
cooling and the observed optical emission.  We note however, that this
scenario has   some  difficulties associated   with the prediction  of
instabilities produced  by the drift of  electrons from the  low field
channel to  the  high   field  boundary  layer, which    would quickly
isotropize  the    beam (\cite{achatz};     \cite{meisenheimer}).  Our
observations  do not currently rule out  the hollow channel model.  We
note that  a  CI  spectrum would be   more  suitable to  describe such
scenario. Observations over  broader frequency range are needed to
test it.

The constant radio-to-optical spectral index in the jet of 3C264 could
be explained with the local reacceleration model (\cite{meisenheimer})
or, alternatively, with the  model  by Heinz \&  Begelman  (1997).  In
both cases, the appearance of the jet is dominated by the fluctuations
of the  magnetic field. The derived dependence  of the  magnetic field
with the distance   from the core of  3C264   ($B\propto D^{-0.8}$) is
smooth, if compared with the expected  trend of an adiabatic expanding
jet ($B\propto D^{[-1,-2]}$).  A similar behavior is also found in the
jet of  M87    (\cite{meisenheimer}).   Such  smooth   dependence   is
consistent with the existence  of mechanisms of field regeneration  or
amplification in the jet (\cite{hughes89}).  For example, regeneration
of the magnetic field could be produced by turbulence in the jet flow,
which in  addition is also  an efficient  mechanism of local  particle
reacceleration (\cite{eilek}).  Another  possible  explanation for the
slow evolution of the magnetic field in 3C264 is that the expansion of
the jet  could be highly influenced  by the properties of the external
medium,  instead of being adiabatic.   In  fact, the morphology of the
radio jet clearly indicates that it  is not expanding freely, and that
the influence of the external  medium might be important. Finally,  as
suggested for M87 by Heinz \&  Begelman (1997), the possibility exists
that the condition  of equipartition,   which  is under all   previous
calculations  but without a strong   justification,  is not valid   in
3C264. 
 
In  summary,  although the mechanism which   maintains or provides the
energy of  the electrons in  the  jet  of 3C264  remains unclear,  the
constancy of the jet radio to optical spectral index suggests that local 
reacceleration or departures from equipartition are
the  most plausible scenarios  for the jet optical emission.

\section{Conclusions}

We  have presented  and  discussed combined EVN  +  MERLIN data of the
radio galaxy 3C264. Detailed images of the radio structure from parsec
to sub-kiloparsec scales   have been obtained, showing  four different
regions  in  the jet:  {\em  i)} the  strong compact  core  and smooth
innermost jet (0 -   10 pc): {\em ii)}  a  well collimated  and narrow
region  (10 -100 pc); {\em iii)}  a region with strong widening, kinks
and filaments, which could be  interpreted as a helical or oscillating
jet  (100 - 300  pc);  {\em iv)} a  faint  and narrow region appearing
after a strong jet deflection (300 - 430 pc). We have shown that there
is  good correspondence between optical  and radio features in the jet
of  3C264.  Moreover, the    radio    to optical spectral  index    is
approximately constant all along the jet. 

The application of  an adiabatic expansion model to  the jet  of 3C264
requires deceleration of the  plasma flow between  100  and 200  pc in
order to compensate  for the expansion of the  jet in this  region and
reproduce the observed   surface  brightness.  However,  the  observed
magnetic field   configuration  and the external   medium, which could
influence the jet expansion, renders this result dubious. 

Possible  mechanisms for  optical emission in   the  jet of 3C264  are
discussed, with synchrotron being  the most plausible.  The integrated
broadband spectrum  can  be better fit  with  a KP 
synchrotron aging model.  We derive an injection spectral index of 0.5
and a  break frequency between radio and  optical wavelengths of $\sim
2\times 10^{14}$  Hz.   However, more data  covering  a wide  range of
frequencies   and  spatial   resolutions  are    necessary  to clearly
discriminate  between the  different   models, and  to   make detailed
studies of the spectral evolution along the jet. 

We have estimated   physical  parameters along the  jet  under minimum
energy considerations. The trend  of the equipartition  magnetic field
with  distance  does  not agree   with   expectations in an  adiabatic
expanding jet.   The  discrepancy could   be  due either to  incorrect
assumptions (geometry of the  magnetic field), or to special  physical
conditions (deviations  from equipartition, existence of amplification
of the magnetic field).    The existence of local   reacceleration or,
alternatively,  departures from the   condition  of equipartition  are
possible scenarios to explain the constant spectral  index and the low
synchrotron cooling in the jet of 3C264. 

The radio structure found in 3C264, that  is, a well collimated narrow
jet which suddenly  widens, is not  unique.  Similar  radio structures
are seen  in  many other FR-I sources.    If the condition  of similar
radio  morphology  could be generalized  to  the condition  of similar
physical properties, then  a large number of  FR-I  jets emitting also
optical synchrotron  radiation should be expected.

\acknowledgments

We  acknowledge Dr.  Antonio Alberdi  and  Dr.  Alan Bridle for  their
valuable comments on the manuscript, and Matteo Murgia for allowing us
to use his code  SYNAGE.  This research was  supported in part by  the
Spanish DGICYT  research  grant PB94-1275.   LL wishes  to acknowledge
support  for   this research  by  the  European  Union under  contract
ERBFMGECT 950012.  The    National Radio Astronomy   Observatory  is a
facility of the National Science Foundation operated under cooperative
agreement  by Associated Universities, Inc.    This work was based  on
observations made with  the NASA/ESA Hubble  Space Telescope, obtained
from the data archive at the Space Telescope Science Institute.

\clearpage

\figcaption[]{Maps of 3C264 at 5.0 GHz with different angular
resolutions. For each map we list the instrument, the Gaussian beam
used for convolution, the lower contour (at 3$\sigma$
of the rms of the image) and the peak of brightness. In all cases contours
increase in powers of 1.5.
(a) EVN; 6 mas $\times$ 3.6 mas in P.A.70$^{\circ}$; 0.6 mJy/beam;
0.126 Jy/beam.
(b) EVN + MERLIN; 9.9 mas $\times$ 6.3 mas in P.A.70$^{\circ}$; 0.3 mJy/beam;
0.137 Jy/beam.
(c) EVN + MERLIN; 18.5 mas $\times$ 15.2 mas in P.A.72$^{\circ}$;
0.3 mJy/beam; 0.155 Jy/beam.
(d) EVN + MERLIN; 23.2 mas $\times$ 20.5 mas in P.A.68$^{\circ}$;
0.3 mJy/beam; 0.160 Jy/beam.
(e) EVN + MERLIN; 35.2 mas $\times$ 31.5 mas in P.A.50$^{\circ}$;
0.3 mJy/beam; 0.165 Jy/beam.
(f) EVN + MERLIN; 62 mas $\times$ 52 mas in P.A.47$^{\circ}$; 0.35 mJy/beam;
0.170 Jy/beam.\label{fig1}}

\figcaption[]{(a) Flux density along the jet ridge line at radio
(solid line) and optical (dashed line) wavelengths. Both brightness scales
have been adjusted for display purposes. (b) Radio to optical spectral index. 
The dotted line represents the average value of the spectral
index along the jet.\label{fig2}}

\figcaption[]{Flux density of the jet of 3C264 vs frequency. 
Straight lines represent independent linear fits for the radio and optical
data. Asterisks correspond to Crane et al. (1993) data.\label{fig3}}

\figcaption[]{Evolution of physical parameters along the jet of
3C264. (a) Deconvolved jet FWHM {\em vs} distance from the core. The dotted
lines represents the aperture of the jet used for the adiabatic model
(see Section 4.1). (b) Minimum pressure at different positions along the jet,
obtained from minimum energy conditions. (c-d) Magnetic field
{\em vs} jet FWHM and distance from the core, respectively. The lines
represent linear fits and the numbers the respective slopes.\label{fig4}}

\figcaption[]{Fits of CI, JP, and KP synchrotron spectral aging models
to the overall spectrum  of the jet of 3C264.\label{fig5}}

\newpage

\clearpage
 
\begin{deluxetable}{ccccc}
\footnotesize
\tablecaption{Jet to Counter-Jet Constraints.\label{tab1}}
\tablewidth{0pt}
\tablehead{
\colhead{Distance} & \colhead{Brightness}&\colhead{R} 
&\colhead{$\beta\cos\theta$} & \colhead{beam} \nl
\colhead{(pc)}   & \colhead{(mJy/beam)}&  &  & \colhead{(mas)} 
} 
\startdata
  4 & 6.25 & $>$9.9  & $>$0.400 & 4.3$\times$4.3 \nl
  6 & 1.90 & $>$3.0  & $>$0.200 & 4.3$\times$4.3 \nl
  8 & 1.86 & $>$3.0  & $>$0.200 & 4.3$\times$4.3 \nl
 10 & 1.25 & $>$2.0  & $>$0.128 & 4.3$\times$4.3 \nl
 20 & 5.55 & $>$18.5 & $>$0.493 &  23$\times$20  \nl
 30 & 1.47 & $>$4.9  & $>$0.286 &  23$\times$20  \nl
 40 & 2.75 & $>$9.2  & $>$0.389 &  23$\times$20  \nl
 50 & 1.58 & $>$5.3  & $>$0.299 &  23$\times$20  \nl
 60 & 1.19 & $>$4.0  & $>$0.251 &  23$\times$20  \nl
 70 & 0.91 & $>$3.0  & $>$0.200 &  23$\times$20  \nl
 80 & 1.80 & $>$6.0  & $>$0.320 &  23$\times$20  \nl
120 & 5.98 & $>$16.6 & $>$0.477 & 108$\times$89  \nl
144 & 8.06 & $>$22.4 & $>$0.519 & 108$\times$89  \nl
160 & 8.76 & $>$24.3 & $>$0.530 & 108$\times$89  \nl
200 & 5.69 & $>$15.8 & $>$0.470 & 108$\times$89  \nl
240 & 5.85 & $>$16.3 & $>$0.475 & 108$\times$89  \nl
320 & 2.78 & $>$7.7  & $>$0.361 & 108$\times$89  \nl
\enddata
\end{deluxetable}

\begin{deluxetable}{ccl}
\footnotesize
\tablecaption{3C264 Jet Flux Density.\label{tab2}}
\tablewidth{0pt}
\tablehead{
\colhead{Frequency} & \colhead{Flux Density} & \colhead{References} \nl
  \colhead{(GHz)}   & \colhead{(mJy)}        &   
}
\startdata
1.66     &  130.0        & MERLIN data, to be published  \nl
5.0      &   78.0        & These data                    \nl
3.83e5   &    0.076      & HST public archive \nl
4.35e5   &    0.062      & BOG97                   \nl
4.46e5   &    0.049      & HST public archive  \nl
5.48e5   &    0.046      & HST public archive  \nl
6.18e5   &    0.093      & Crane et al. 1993  \nl
8.82e5   &    0.059      & Crane et al. 1993  \nl
242e6    &   $<$0.0005   & Prieto et al. 1996  \nl
\enddata
\end{deluxetable}

\begin{deluxetable}{ccccccc}
\footnotesize
\tablecaption{Physical Parameters along the Jet of 3C264.\label{tab3}}
\tablewidth{0pt}
\tablehead{
\colhead{Distance} & \colhead{FWHM} & \colhead{Brightness} & \colhead{B$_{me}$} & \colhead{u$_{me}$} & \colhead{P$_{eq}$}& \colhead{t$_{\nu_{b}}$} \nl
  \colhead{(pc)}  & \colhead{(pc)} &\colhead{(Jy/asec$^2$)}
&\colhead{($\eta^{-2/7}$ mG)}& \colhead{($\eta^{-4/7}$ erg/cm$^{3}$)} 
& \colhead{($\eta^{-4/7}$ dyn/$cm^{2}$)} & \colhead{($\eta^{6/7}$ yr)}
} 
\startdata
  0&  0.29 &52430 & 65.45 & 3.98$\times 10^{-4}$ & 2.46$\times 10^{-4}$ &0.14\nl
  2&  0.70 & 3242 & 22.97 & 4.90$\times 10^{-5}$ & 3.04$\times 10^{-5}$ &0.74\nl
  4&  0.91 &  917 & 14.86 & 2.05$\times 10^{-5}$ & 1.27$\times 10^{-5}$ &1.42\nl
  8&  1.20 &  224 &  9.18 & 7.81$\times 10^{-6}$ & 4.84$\times 10^{-6}$ &2.92\nl
 20&  4.18 & 42.2 &  4.00 & 1.48$\times 10^{-6}$ & 9.15$\times 10^{-7}$ &10.2\nl
 30&  9.43 & 5.50 &  1.77 & 2.90$\times 10^{-7}$ & 1.80$\times 10^{-7}$ &35  \nl
 40&  7.00 & 9.34 &  2.24 & 4.64$\times 10^{-7}$ & 2.88$\times 10^{-7}$ &24  \nl
 50&  4.64 & 9.20 &  2.51 & 5.82$\times 10^{-7}$ & 3.61$\times 10^{-7}$ &20  \nl
 60&  4.21 & 7.48 &  2.43 & 5.47$\times 10^{-7}$ & 3.39$\times 10^{-7}$ &21  \nl
 70& 10.89 & 3.23 &  1.46 & 1.96$\times 10^{-7}$ & 1.22$\times 10^{-7}$ &46  \nl
 80&  9.64 & 6.67 &  1.85 & 3.19$\times 10^{-7}$ & 1.98$\times 10^{-7}$ &32  \nl
120& 24.20 & 1.38 &  0.91 & 7.67$\times 10^{-8}$ & 4.75$\times 10^{-8}$ &94  \nl
144& 37.24 & 1.40 &  0.81 & 6.04$\times 10^{-8}$ & 3.75$\times 10^{-8}$ &112 \nl
160& 52.42 & 1.29 &  0.72 & 4.75$\times 10^{-8}$ & 2.95$\times 10^{-8}$ &134 \nl
200& 57.96 & 0.81 &  0.61 & 3.43$\times 10^{-8}$ & 2.13$\times 10^{-8}$ &171 \nl
240& 65.22 & 0.80 &  0.59 & 3.19$\times 10^{-8}$ & 1.98$\times 10^{-8}$ &181 \nl
320& 42.8  & 0.45 &  0.56 & 2.91$\times 10^{-8}$ & 1.81$\times 10^{-8}$ &193 \nl
\enddata
\end{deluxetable}

\clearpage

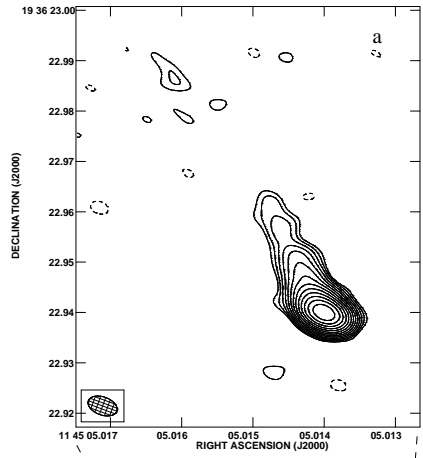
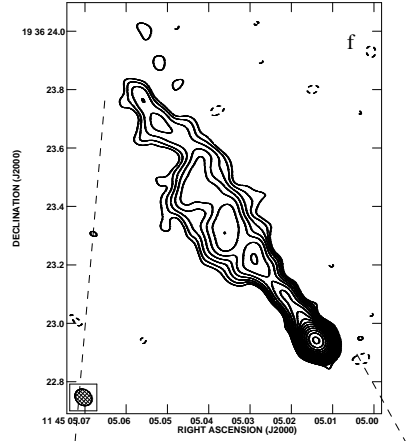
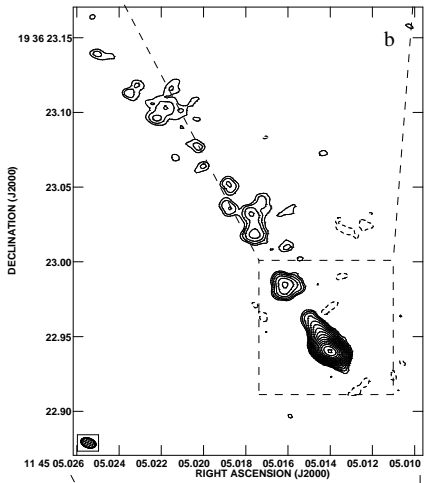
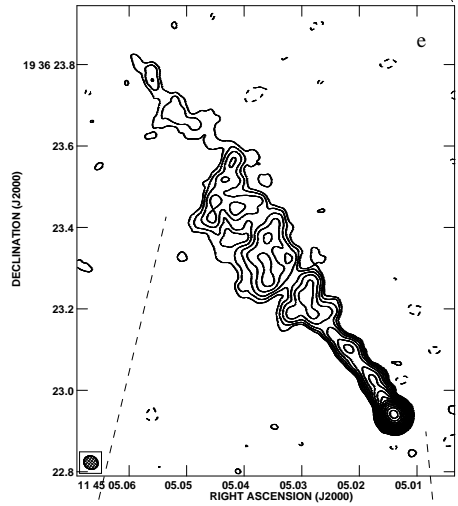
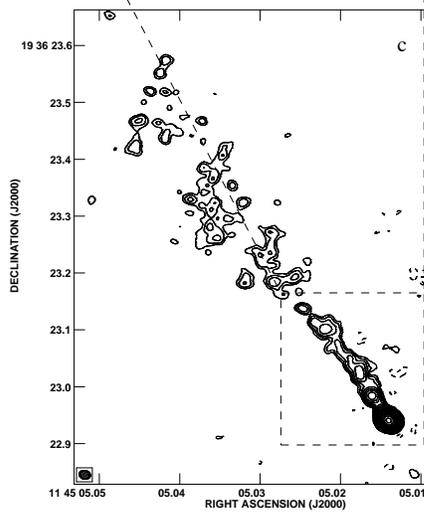
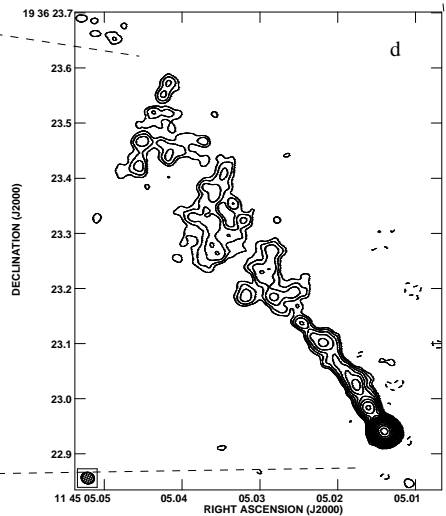
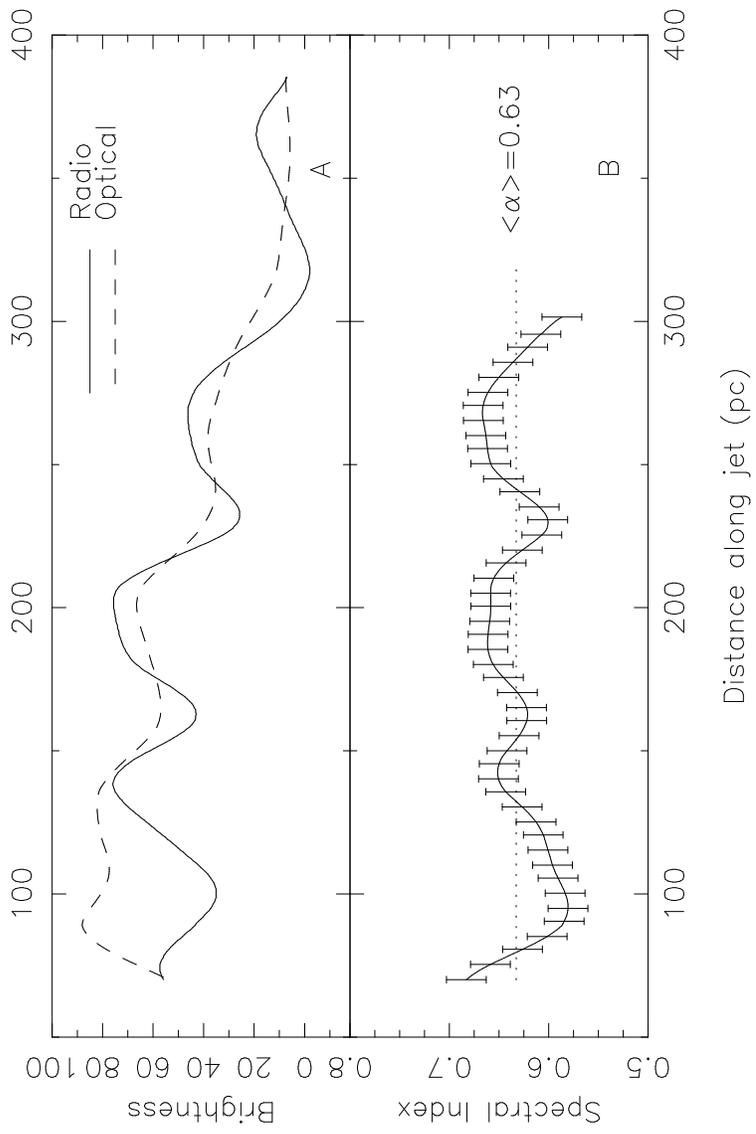
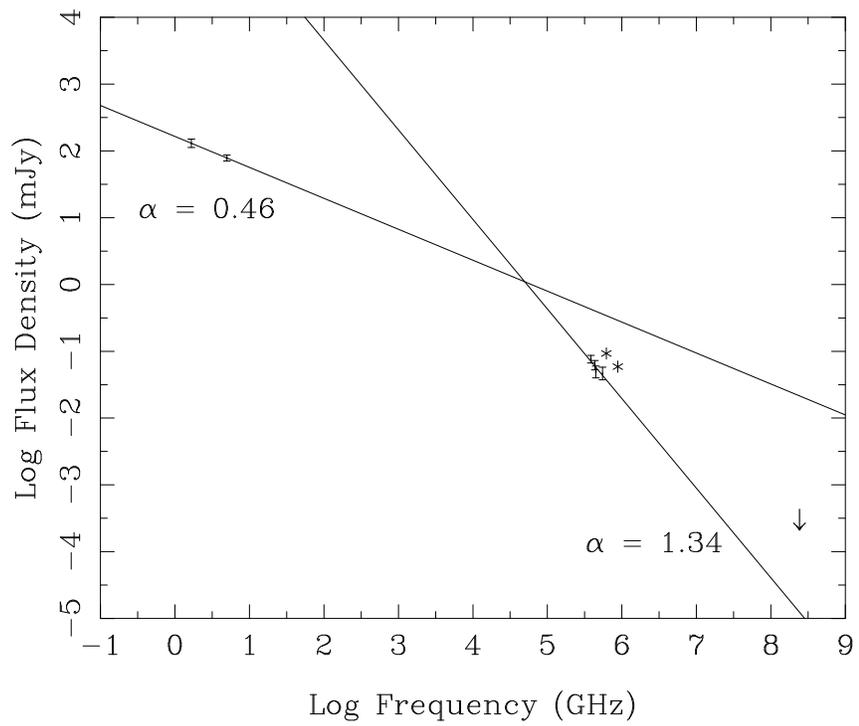
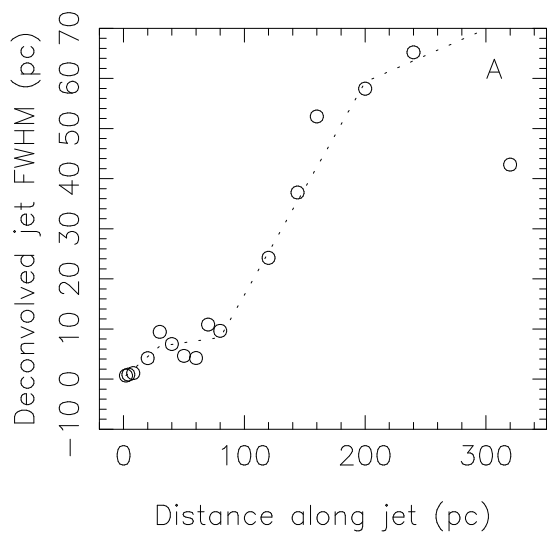
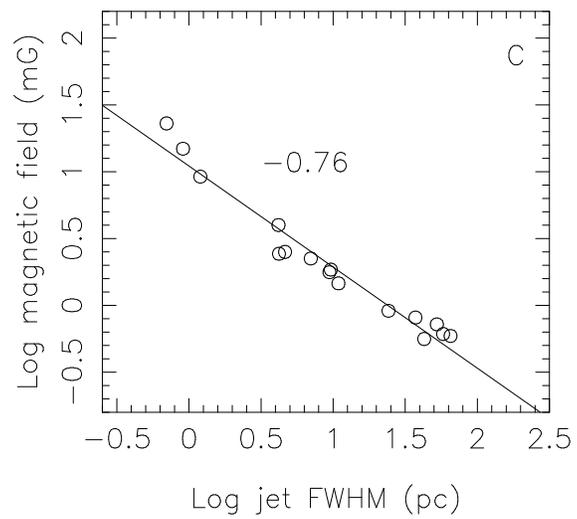
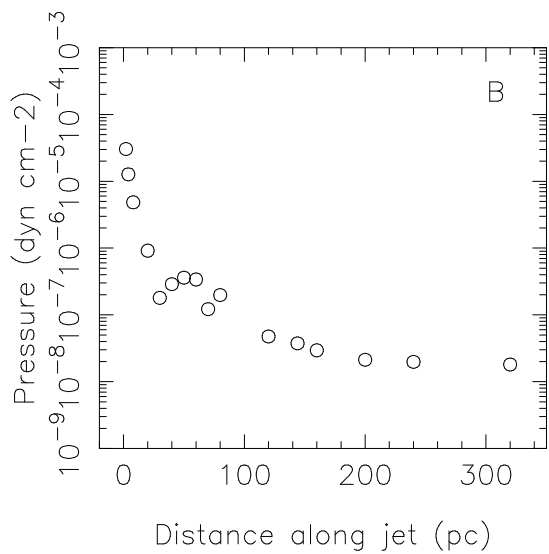
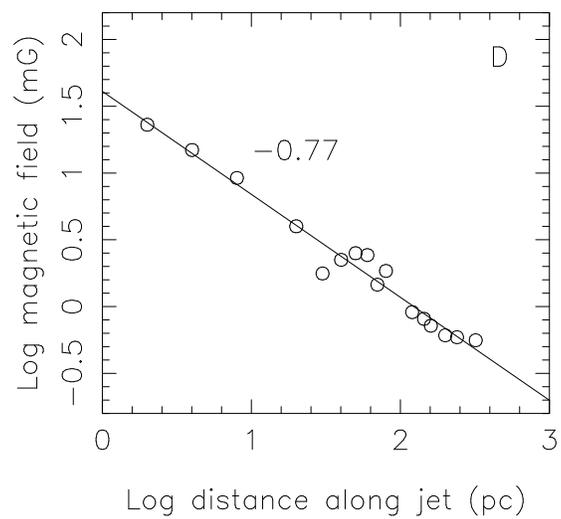
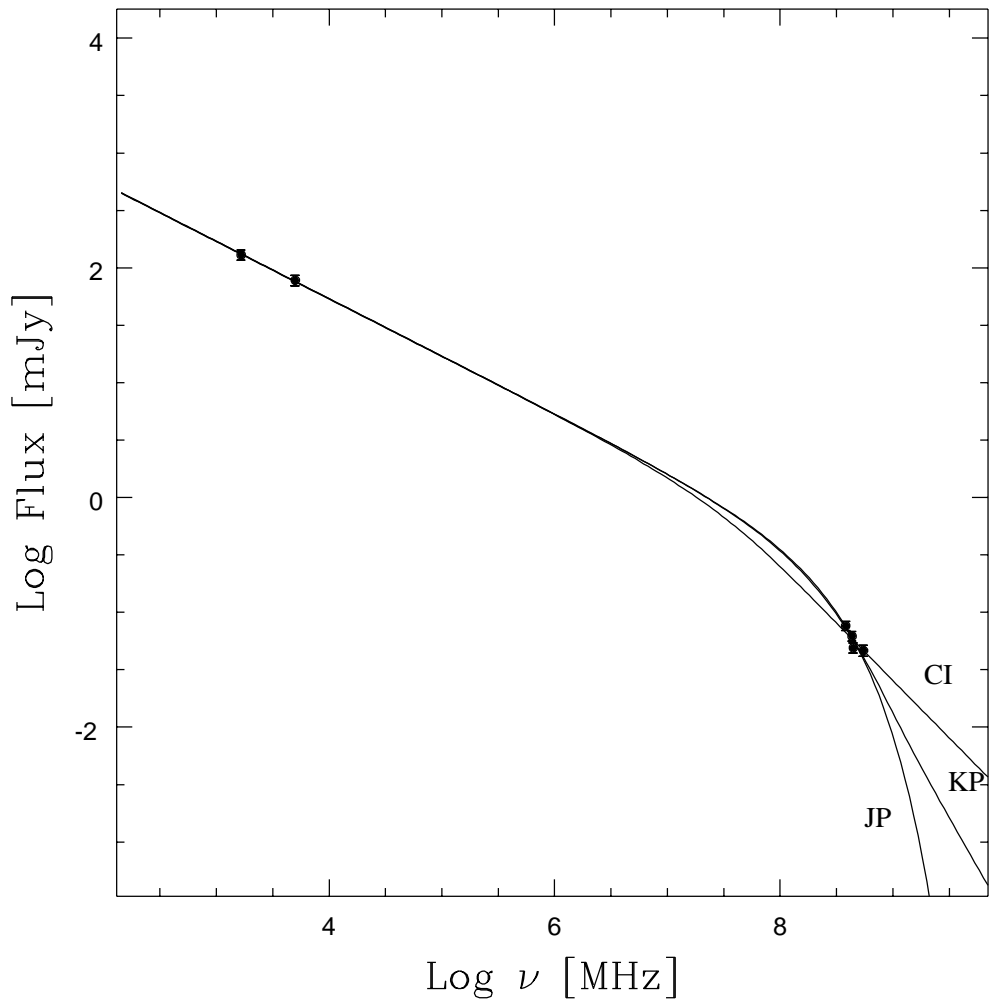
\end{document}